\documentclass[12pt]{iopart}
\usepackage[dvips]{color,graphicx}

\begin{document}

\title[Dynamics in binary cluster crystals]{Dynamics in binary cluster crystals}

\author{Manuel Camargo$^{1,2}$, Angel J Moreno$^{3}$, and Christos N
Likos$^{4,1}$}
\address{$^1$ Institute of Theoretical Physics, Heinrich Heine University of
D{\"u}sseldorf, Universit{\"a}tsstra{\ss}e 1, D-40225 D{\"u}sseldorf, Germany}
\address{$^2$ Direcci\'on Nacional de Investigaciones, Universidad Antonio
Nari\~no, Kra 3 Este 47a-15 Bogot\'a, Colombia}
\address{$^3$ Centro de F\'isica de Materiales (CSIC,UPV/EHU) and Materials
Physics Center MPC, Paseo Manuel de Lardizabal 5, E-20018 San Sebasti\'an,
Spain}
\address{$^4$ Faculty of Physics, University of Vienna, Boltzmanngasse 5,
A-1090 Vienna, Austria}
\eads{\mailto{camargo@thphy.uni-duesseldorf.de}, \mailto{wabmosea@ehu.es}, and
\mailto{christos.likos@univie.ac.at}}

\begin{abstract}
As a result of the application of coarse-graining procedures to describe complex
fluids, the study of systems consisting of particles interacting through
bounded, repulsive pair potentials has become of increasing interest in the last
years. A well known example is the so-called Generalized Exponential Model
(GEM-$m$), for which the interaction between particles is described by the
potential $v(r)=\epsilon\exp[-(r/\sigma)^m]$. Interactions with $m > 2$ lead to
the formation of a novel phase of soft matter consisting of cluster crystals.
Recent studies on the phase behavior of binary mixtures of GEM-$m$ particles
have provided evidence for the formation of novel kinds of alloys, depending on
the cross interactions between the two species. This work aims to study the
dynamic behavior of such  binary mixtures by means of extensive molecular
dynamics simulations, and in particular to investigate the effect of the
addition of non-clustering particles on the dynamic scenario of one-component
cluster crystals. Analogies and differences with the 
one-component case are revealed and discussed by analyzing self- and collective
dynamic correlators.

\end{abstract}

\pacs{61.20.Ja, 64.70.Pf, 82.70.-y}\vspace{2pc}
\noindent{\it Keywords}: soft matter, diffusion, cluster crystal, hopping

\maketitle

\section{Introduction}

Coarse-graining procedures provide a general way to represent mesoscopic
aggregates of several architectures (e.g., linear and hyperbranched polymers,
micelles, microgels, etc.) as particles interacting via an effective, isotropic
potential $V_{\rm eff}(r)$. When the center of mass of such aggregates can
coincide without violating excluded volume interactions, $V_{\rm eff}(r)$ turns
out to be ultrasoft and bounded \cite{louis:prl:2000,bolhuis:jcp:2001,
likos:jcp:2002}. These characteristics may lead to cluster formation providing
$V_{\rm eff}(r)$ belongs to the so-called $Q^\pm$-class, i.e. its Fourier
transform (FT) $\hat V_{\rm eff}(q)$ displays an oscillatory decay around zero
(potentials with positive FT are referred as $Q^+$ potentials) 
\cite{likos:pre:2001,glaser:epl:2007}.  Recent numerical investigations of
amphiphilic dendrimers and of ring polymers with different degree of knotedness
demonstrate that such macromolecules are candidates for experimental
realizations
of the former clustering scenario, since their effective interactions are
indeed of the $Q^\pm$-class \cite{mladek:prl:2008,lenz:sm:2009, narros:sm:2010}.
A particular theoretical realization of $Q^\pm$ potentials is given by the 
generalized exponential model (GEM-$m$), for which the interaction is described
by the potential
\begin{equation}
\label{eq:gem}
v(r) = \epsilon\exp
\left[-\left(\frac{r}{\sigma}\right)^m\right]\hspace{2cm}(m>2),
\end{equation} 
where $r$ is the interparticle distance, $\sigma$ is a measure of the particle
size, and $\epsilon$ is the interaction strength. As density increases, the
GEM-$m$ system shows, for $m>2$, a first-order transition
from a fluid of clusters to a cluster bcc-crystal above the triple point
temperature, and a subsequent structural phase transition from a cluster bcc to
a cluster fcc crystal. This occurs at arbitrarily high temperatures
\cite{likos:jcp:2007, mladek.prl.2008}, contrary to the case of $Q^+$-systems,
which exhibit reentrant crystallization  in the temperature-density plane
\cite{lang:jphys:2000}. A remarkable feature  of the  cluster fcc
crystals is that upon increasing the density the lattice constant approaches a
constant value  $a_{\rm fcc}=2\pi\sqrt{3}/q^*$, with   $q^*$ the wave vector at
which $\hat v(q)$ takes its  absolute minimum. An obvious consequence of the
density-independent value of $a_{\rm fcc}$ is that the population of each
lattice site scales proportionally with density. These properties are rather
different from  usual crystallization features in colloidal systems. Moreover,
these differences are not restricted to the structural properties but also to
the dynamical ones. Regarding the slow dynamics in such cluster crystals,
incessant hopping between clusters has been revealed, which fully changes the
initial identity of the clusters without altering the lattice structure. Another
peculiar dynamic feature of these systems is a decoupling between self- and
collective out-of-lattice correlations \cite{moreno:prl:2007,likos:cpc:2008}. 

Recently, the phase behavior of mixtures containing a non-clustering component
($m=2$) and a clustering component ($m=4$) has been investigated within the
framework of the Density Functional Theory (DFT) \cite{sarah:epl:2009,
sarah:jcp:2009}. For the case of non-demixing systems, evidence  was found for
the formation of novel kinds of alloys, i.e. mixed cluster crystals
\cite{sarah:jcp:2009}. Though static features of these mixtures have been
studied in detail, to the best of our knowledge no information has been reported
on the corresponding dynamical aspects. This article aims to shed light
on such aspects. By means of extensive molecular dynamics (MD) simulations, the
findings displayed by one-component cluster crystals
\cite{moreno:prl:2007,likos:cpc:2008} are extended to mixtures of clustering and
non-clustering particles. Analogies and differences with the one-component case
are revealed and discussed by analyzing self- and collective dynamic
correlators. The article is organized as follows: Model and simulation details
are given in Section 2. Static and dynamic features are presented and discussed
in Section 3. Conclusions are given in Section 4.

\section{Model and Simulation}

As a previous step, the stability limits of the homogeneous fluid phase
(coexistence and $\lambda$ lines) were determined for representative examples of
the former class of binary mixtures. These limits were obtained by solving the
Ornstein-Zernike (OZ) equation within the mean field approximation (MFA)
\cite{hansen:tsl:2006}. Based on such results, the specific systems to be
simulated were determined and thus the parameters defining the mixture
(relative particle size, mixture  compositions, temperature and density, self-
and cross-interactions) were chosen in such a way as 
to avoid the region of demixing in
the phase diagram. Because of the high total densities to be considered, the
former parameters were also selected with the goal of reducing computational
expense.

The investigated system was a mixture of big GEM-8 particles and small GEM-2
(gaussian) particles. This mixture contains a non-clustering component (GEM-2,
in the following referred as $A$-particles) and a cluster-forming component
(GEM-8, in the following referred as $B$-particles). The dynamic features of
this system were investigated at fixed temperature $T$ for a broad range
of densities $\rho=\rho_A+\rho_B$ and several relative compositions $x =
\rho_B/\rho$. The former densities are defined as 
$\rho_\alpha = N_\alpha/V$, $N_\alpha$ being the number of particles of the 
species $\alpha$ and $V$ the total volume. In what follows, we choose units in 
which Boltzmann's constant $k_{\rm B}$ has the value $k_{\rm B}=1$. 
The cross-interaction was chosen to 
be a GEM-4 potential, which also displays cluster formation. In this way, 
the following interaction parameters were adopted for the potential \eref{eq:gem}: 
$\epsilon_{\alpha\beta}=\epsilon=1.0$, $\sigma_{AA}=0.3\sigma$,
$\sigma_{AB}=0.6\sigma$, $\sigma_{BB}=\sigma=1.0$, and $m_{AA}=2$, $m_{AB}=4$,
$m_{BB}=8$. With these parameters, a common cutoff range $R_c=1.5\sigma$ was
introduced in the simulation for all the interactions.

MD simulations were performed in a cubic box with periodic boundary conditions.
The equations of motion were integrated in the velocity Verlet scheme
\cite{frenkelbook}, with a time step $\Delta t/\tau$ ranging from 0.001 to
0.005, where   $\tau=\sqrt{m\sigma^2/\epsilon}$ and a common mass $m=1$ was
used for all the particles. The size of the simulation box, $L_{\rm box}$, was
typically of 7 or 8 times the lattice constant $a_{\rm fcc}$ of the considered
mixture. The value of $a_{\rm fcc}$ was obtained from DFT as explained in
\cite{sarah:epl:2009, sarah:jcp:2009}. In the following, density, time,
distance, and wave vector will be given, respectively, in units of
$\sigma^{-3}$, $\tau$, $\sigma$, and $\sigma^{-1}$. Initial configurations were
generated by placing the particles  uniformly at the sites of the fcc lattice
defined by $a_{\rm fcc}$. An equilibration run was performed in which the system
was thermalized at temperature $T = 0.30$ by periodic velocity rescaling.
Typical equilibration times ranged from $2\times10^5$ to $2\times10^6$ time
steps according to the studied density. After reaching equilibrium, manifested
by the absence of any
drift in internal energy and pressure, a production
run was performed in the microcanonical ensemble, at different densities, for
compositions $x=$ 0.65, 0.80 and 0.95. Typical production runs ranged from
$10^6$ to $10^8$ time steps.

\section{Results and Discussion}

\subsection{ Phase diagram and static structure}

\begin{figure}[!t]
\begin{center}
\includegraphics*[width=0.45\textwidth]{./Fig1a_Phase.eps}\hspace{
0.1\textwidth}
\includegraphics*[width=0.30\textwidth]{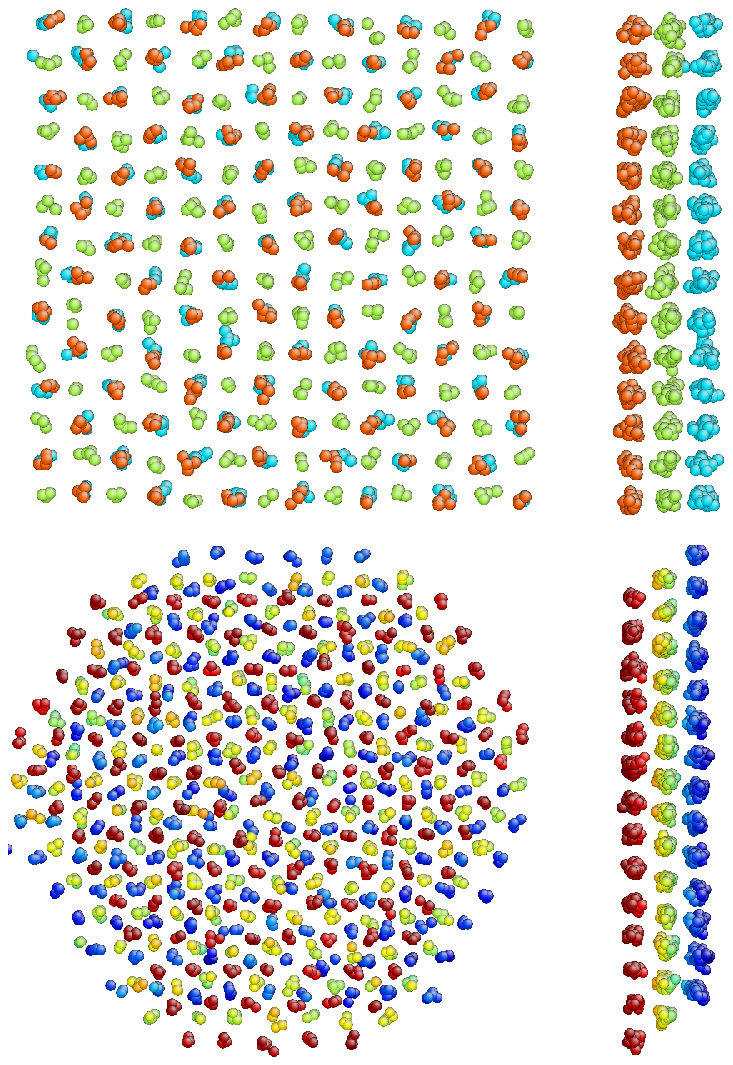}
\end{center}
\caption{Left: A segment,
at the high-$x$/moderate $\rho$ region, of the phase diagram of the
studied ultrasoft mixture. 
Empty and filled symbols indicate the simulated state points in
the $(\rho,x)$-plane, at both sides of the melting line. The latter is displayed
as a dashed guideline for the eyes. It is estimated from the emergence of Bragg
peaks in the partial static structure factors (see below). The solid line is the
$\lambda$-line, obtained from the OZ-MFA equations. Right: Slabs of the
simulation box for $x=0.65$ and $\rho=3.50$, parallel to the $(001)$-plane (top)
 and to the $(111)$-plane (bottom). For the sake of clarity, only GEM-8
particles are displayed, in both frontal and lateral views, with a size much
smaller than the real $\sigma$. Particles in different layers are represented
with different colors.}
\label{Fig:phase}
\end{figure}

The partial structure factors $S_{\alpha\beta}(q)$ ($\alpha,\beta \in \{A,B\}$)
are related to the propensity of a fluid to sustain spontaneous density
fluctuations of wave vector $q$. They are defined as 

\begin{equation}
S_{\alpha\beta}(q) = \frac{\langle
\rho_{\alpha}({\bf q},0)\rho_{\beta}(-{\bf
q},0)\rangle}{\sqrt{N_{\alpha}N_{\beta}}} 
\label{sab:eq}
\end{equation}
with $\rho_{\alpha}({\bf q},t) = \sum_{j=1}^{N_\alpha} \exp[i{\bf q}\cdot {\bf
r}^\alpha_j (t)]$ and the sum is performed over the coordinates ${\bf
r}^\alpha_j$ of all particles belonging to the species 
$\alpha$.\footnote{For uniform phases, the statistical average
$\langle\cdots\rangle$
in the right hand side of eq.~(\ref{sab:eq}) above renders its left hand side
a function of $q = |{\bf q}|$ only. For the crystalline phases to be considered
in what follows, we have performed an additional rotational average over
symmetry-related values of the reciprocal vectors, so that the resulting
quantities (static and time-dependent correlation functions) are shown as
functions of $q$ only.}
According to
the OZ equations they take, in the homogeneous phase, the generic form
$S_{\alpha\beta}(q)=N_{\alpha\beta}(q)/D(q)$ 
where, within the MFA, the denominator depends  on the Fourier transform of the
interaction potentials according to 
\begin{equation}
D(q)= 1 + \frac{\rho}{T} \hat U_s(q) + \left(\frac{\rho}{T}\right)^2x(1-x)
\hat U_d(q),
\label{Eq:Doq}
\end{equation}
where $\hat U_s(q)= (1-x) \hat v_{AA}(q) + x \hat v_{BB}(q)$ and $\hat U_d(q)=
\hat v_{AB}(q)\hat v_{BB}(q)-\hat v_{AB}^2(q)$ \cite{sarah:epl:2009}. Due to
their generic form, the partial structure factors will diverge for all
those conditions where $D(q)=0$, and therefore the homogeneous fluid will be
unstable. If the divergence takes place at $q=0$, demixing occurs and the
spinodal line denotes the locus of state points where $D(0)=0$.  On the other
hand, a divergence at some $q^*>0$ signals an instability in the fluid (which is
referred as Kirkwood instability) with respect to a periodic modulation of the
density. The locus of state points where this instability takes place 
corresponds to the so-called $\lambda$-line \cite{sarah:jcp:2009}. 

For the system at hand, the last scenario is indeed the case, where the
instability  dominating the phase behavior is mainly due to the  $Q^\pm$-nature
of the species $B$. \Fref{Fig:phase} displays in the $(x,\rho)$-plane the
corresponding $\lambda$-line estimated from MFA. Since this line is related to
the limit of stability for the homogeneous fluid phase, it is expected  that
close to it the system will tend to crystallize into solids whose lattice
constant will be dictated by $q^*$ (see Table \ref{tab:fcc}). By visual
inspection of
the configuration snapshots, state points for which the initial crystal
structure was stable were discriminated from those for which the latter melted.
The melting line estimated in this way was consistent with the analysis of Bragg
peaks in the partial structure factors (see below). No transition to other
crystalline structures was observed during the simulations, confirming the fcc
structure as the underlying lattice in the investigated crystalline states. The
typical snapshots of \Fref{Fig:phase} indeed exhibit the expected ABC staking of
the (111)-planes. It should be noted that the apparent contradiction of
having, in \Fref{Fig:phase}, a fluid phase that is stable {\it beyond} the
$\lambda$-line is an artefact arising from applying the MFA in calculating
the latter. The MFA becomes increasingly accurate at $T$ grows, whereas
we are working here at $T=0.3$. Nevertheless, this has no consequences
in what follows, since we are employing the MFA $\lambda$-line merely as
an indicator of the region in which alloy formation (crystallization) is
expected, without basing any further quantitative predictions on its
precise location. 

\Table{\label{tab:fcc}
For each investigated composition $x$, 
density $\rho^*$ for the
$\lambda$-line, parameters defining the fcc lattice ($D(q^*)=0$, $a_{\rm
fcc}=2\pi\sqrt{3}/q^*$, see \cite{sarah:epl:2009}) and box size $L_{\rm box}$
used in the MD.}
\br
$x$  & $\rho^*$       & $q^*$  & $a_{\rm fcc}$ & $L_{\rm box}$ \\
\mr
0.65 & 1.68           & 5.64   & 1.9296  & 13.5070 \\
0.80 & 1.46           & 5.74   & 1.8960  & 15.1677 \\
0.95 & 1.29           & 5.84   & 1.8635  & 14.9079 \\
1.00 & 1.25           & 5.86   & 1.8571  & 14.8571 \\ 
\br
\endTable

\begin{figure}[!t]
\begin{center}
\includegraphics*[width=0.93\textwidth]{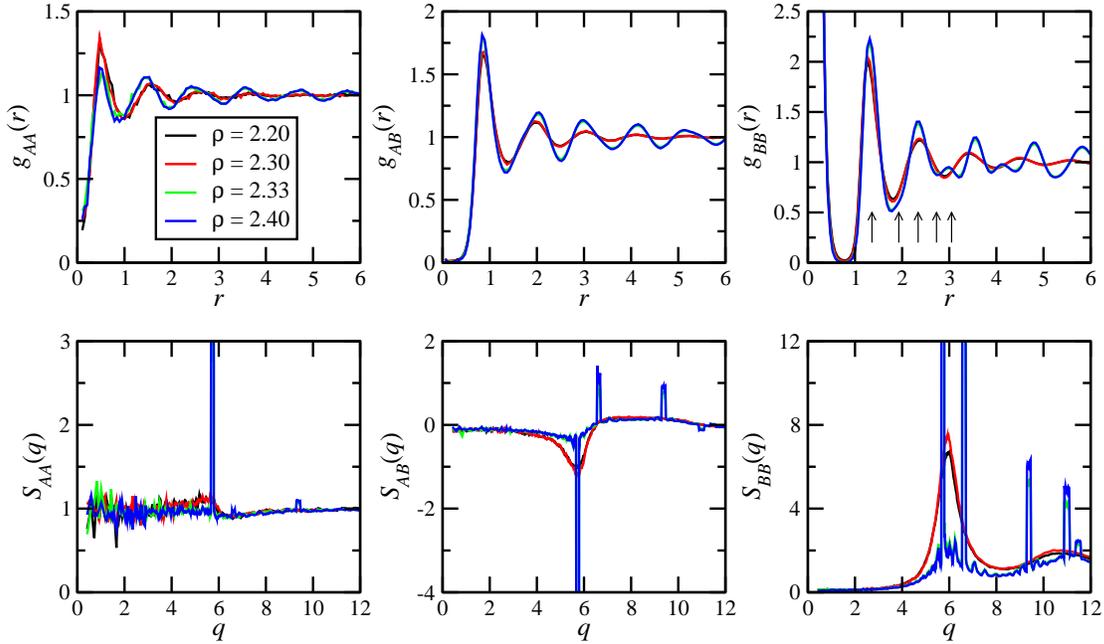}
\end{center}
\caption{Radial distribution functions (top) and structure factors (bottom) for
$x=0.80$ and different densities. The arrows in the plot of $g_{BB}(r)$ indicate
the position of the $k$th-neighbor in a fcc lattice, i.e. $d_{k}=\sqrt{k}d_{\rm
nn}=\sqrt{k/2}~a_{\rm fcc}$. The values of the selected densities are the same
in all panels (see legend for $g_{AA}(r)$).}
\label{Fig:grsq}
\end{figure}

More detailed structural information can be gained by considering the partial
radial distribution functions $g_{\alpha\beta}(r)$, which measure the
probability to find a pair of particles, of species $\alpha$ and $\beta$, at a
mutual distance $r$. In \fref{Fig:grsq}  these quantities are displayed for
$x=0.80$ and different densities. As expected, $g_{BB}(r)$ exhibits a large peak
at $r=0$ (namely $g_{BB}(r\to0)\sim 30$), indicating the formation of clusters,
whose size extends up to $d_{\rm c} \sim0.75$. The second peak at $r\sim 1.34$
corresponds to the nearest-neighbor distance in the fcc crystal, $d_{\rm
nn}=a_{\rm fcc}/\sqrt{2}$. The radial distribution function for distinct pairs,
$g_{AB}(r)$, displays rather different features. The correlation hole at $r\to
0$ indicates that the small $A$-particles avoid the lattice sites and tend to
localize in the interstitials between $B$-clusters. Concerning correlations
between pairs of $A$-particles, $g_{AA}(r)$ displays finite small values ($\sim
0.2$) at $r \to 0$.  This means that for the considered conditions, no
appreciable clustering of the species $A$ is induced by the clustering of the
species $B$.

On increasing the density, a significant structural change occurs in the system,
which is evidenced by the emergence of well-defined peaks of all
$g_{\alpha\beta}(r)$ at large distances (see top panels of \fref{Fig:grsq}).
This change is better reflected in the corresponding partial structure factors
$S_{\alpha\beta}(q)$. As shown in bottom panels for the largest densities, the
emergence of Bragg peaks in $S_{BB}(q)$ indicates a rather ordered structure,
which still retains some characteristics of the fluid phase.  On the other hand,
the nearly flat structure of $S_{AA}(q)$ clearly suggests the presence of a,
nearly ideal, fluid phase of $A$-particles. This fluid phase is immersed in the
crystalline matrix of clusters of $B$-particles. Strictly speaking, the
$A$-particles form a very weakly modulated fluid, so that the mixture of
localized $B$-particles and delocalized $A$-particles forms a sublattice melt
phase. The one-particle density of the $A$-particles, however, cannot be
strictly uniform, since the crystallized $B$-particles act on the $A$-species
as a periodically modulated external potential.
This structural scenario has
its dynamic counterpart, as will be shown below.

The occupation number $n$ of a given lattice site, i.e., the number of
$B$-particles in the associated cluster, is obviously an integer number. The
\textit{average} occupation is in general non-integer, as a result of (at least
two) significant contributions of distinct values of $n$ all over the lattice.
For example, in the crystalline system of composition $x = 0.80$ and density
$\rho = 2.40$ around the $65\%$ and $25\%$ of the clusters are composed
respectively by three and four $B$-particles. This percentages were estimated by
counting the coordination number of each particle for $r < d_{\rm c}\sim0.75$.
On passing, it was also estimated by using the same criterion that around 
$90\%$ of the $A$-particles are isolated from particles of the same species. The
former results reflect that the lattice is full of defects which might be
expected to break its stability. As discussed in Ref. \cite{likos:cpc:2008} for
the one-component cluster crystal, the stability of the lattice is actually
maintained by incessant hopping of the $B$-particles between distinct clusters.
This is also the case in the mixtures considered in the present work, as
discussed in the next subsection.

\subsection{Dynamics in real space}

\begin{figure}[!t]
\begin{center}
\includegraphics*[width=0.93\textwidth]{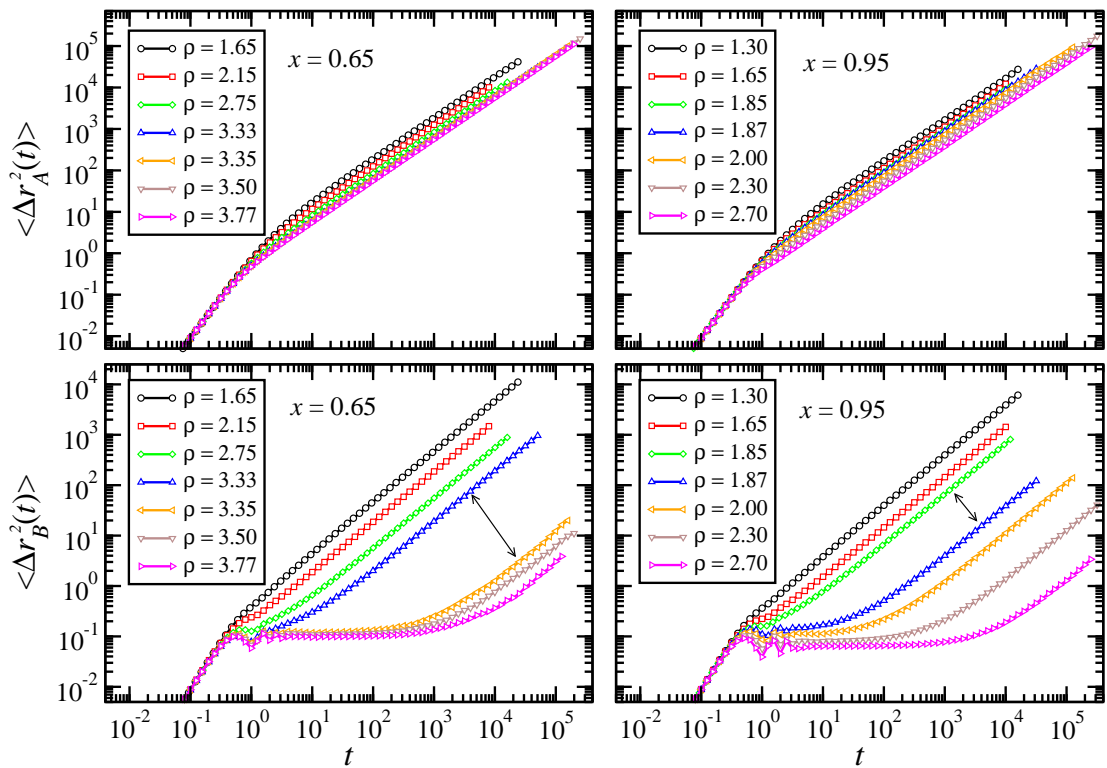}
\end{center}
\caption{MSD of both species for $x=0.65$ (left panels) and $x = 0.95$ (right
panels). The arrows in bottom panels separate fluid states for the $B$-particles
from crystal clusters (see also \Fref{Fig:phase}).}
\label{Fig:msd}
\end{figure}

A first step to gain some insight into the transport properties of the system is
provided by the mean squared displacement  $\left\langle \Delta r_\alpha^2(t)
\right\rangle$ (MSD). This quantity is shown in \fref{Fig:msd} for different
densities and relative compositions $x=0.65$ and $x=0.95$. In the case of the
$A$-particles a rapid crossover is observed, for all the investigated densities,
between short-time ballistic motion ($\langle \Delta r_\alpha^2(t) \rangle
\propto  t^2$) and long-time diffusion ($\langle \Delta r_\alpha^2(t) \rangle
\propto  t$). This is the usual dynamic behavior of a highly mobile fluid
phase. Thus, as anticipated by the static correlations (see above), the mixture
reaches a phase in which the species $A$ shows a fluid-like behavior in a
crystalline matrix of clusters of $B$-particles.

The MSD of the $B$-particles shows a similar simple behavior in the fluid
phase. However, an abrupt dynamic change is revealed by crossing the
crystallization point. Thus, an intermediate plateau regime arises between the
ballistic and diffusive limits, extending over longer time intervals as density
increases. This regime is associated with the temporary trapping of the
$B$-particles within the clusters \cite{moreno:prl:2007,likos:cpc:2008}. The
large oscillations at the beginning of the plateau regime are the signature of
strong intracluster vibrational motion. Such short-time strong oscillations are
an artifact of the simulated \textit{Newtonian} dynamics. They are expected to
be strongly damped under more realistic Brownian dynamics (BD); it has been
shown that they even vanish in Monte Carlo (MC) simulations
\cite{coslovich:arXiv:2010}. Finally, at long times, the MSD displays a
crossover to diffusive behavior. This scenario, which has been previously
observed in the cluster crystal phase of the pure GEM-8 system
\cite{moreno:prl:2007,likos:cpc:2008}, does not originate from the contribution
to the MSD of rare events involving a few diffusing $B$-particles. As
anticipated above, it is the result of incessant hopping of $B$-particles
between neighboring lattice sites. Thus, the hopping mechanism fully changes
the initial identity of each cluster without altering the 
overall crystalline lattice
structure \cite{moreno:prl:2007,likos:cpc:2008}.

\begin{figure}[!t]
\begin{center}
\includegraphics*[width=0.93\textwidth]{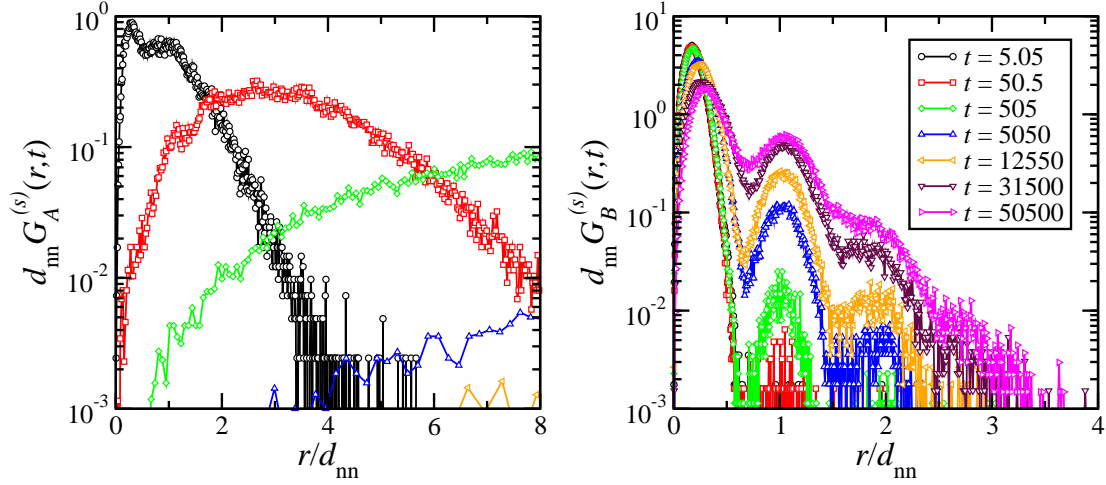}
\end{center}
\caption{Van Hove self-correlation function at different times for $x=0.80$ and
$\rho=3.05$ (left: $A$-particles, right: $B$-particles). Distances are rescaled
by $d_{\rm nn}$ (see text).}
\label{Fig:vhrt}
\end{figure}

This hopping mechanism is clearly reflected in the van Hove self-correlation,
$G^{\rm (s)}_\alpha(r,t)$, of the $B$-particles. \Fref{Fig:vhrt} shows a
representative example for composition $x=0.80$ and density $\rho = 3.05$.  
A sequence of well-defined peaks is present in $G^{(s)}_B(r,t)$, corresponding
to different distances between lattice sites. With increasing time, the first
peak in $G^{\rm (s)}_B(r,t)$ decreases while peaks located at larger distances
grow progressively, corresponding to particles moving away from their original
home clusters. By simple integration of $G^{(s)}_B(r,t)$ from its first minimum
to $r \rightarrow \infty$, it is found that only the 30\% of the $B$-particles
are located at $t \sim 50000$ in their initial home cluster. As expected from
the observations in the MSD (see above) the van Hove self-function for the
non-clustering species $A$ displays a more simple behavior. For $t \sim 50$ the
initial peak at $r \approx 0.3$ has already vanished. At that time scale almost
all the $A$-particles have left their initial interstitial positions and $G^{\rm
(s)}_A(r,t)$ exhibits simple Gaussian behavior.

\begin{figure}[!t]
\begin{center}
\includegraphics*[width=0.93\textwidth]{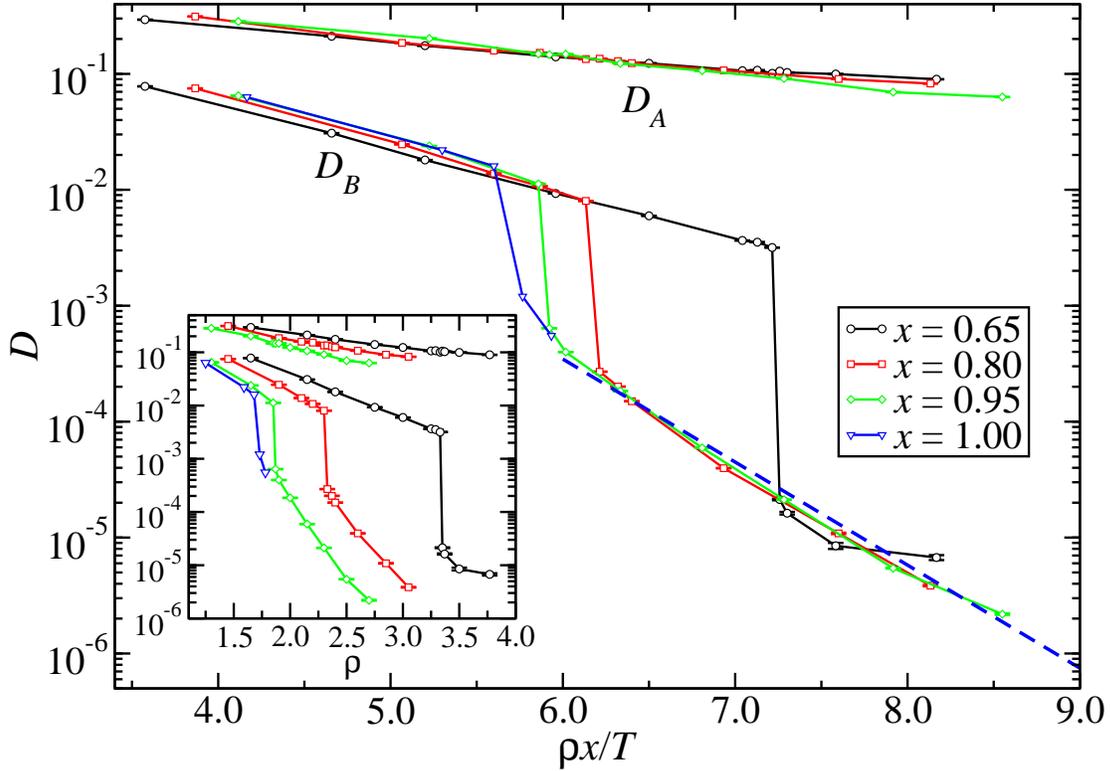}
\end{center}
\caption{Diffusivities of the $A$- and $B$-particles as a function of the
density (inset) and of the combined variable $\rho x/T$ (main panel). Solid
lines are guides for the eyes. The dashed line in the main panel indicates
approximate Arrhenius behavior $D_B \sim \exp[-2.3\rho_B/T]$ (see
text).}
\label{Fig:diff}
\end{figure}

We can estimate the diffusivity of each species from the relation
$D_\alpha=\lim_{t\to\infty}\left\langle \Delta r_\alpha^2(t) \right\rangle/6t$.
\Fref{Fig:diff} shows the obtained results as a function of the quantity
$\rho_B/T =\rho x/T$. For each composition $x$, an abrupt drop in the
diffusivity
of the $B$-particles is observed in a narrow range of density. This drop
reflects the transition of the species $B$ from the fluid to the cluster
crystal, and it is indeed expected from the previous observations in the MSD
(see \Fref{Fig:msd}). The approximate scaling behavior for $D_B$ in the cluster
crystal phase generalizes the observation for the pure GEM-8 system
\cite{moreno:prl:2007,likos:cpc:2008}. The latter was rationalized as the result
of activated hopping motion between minima of the local potential energy, which
are placed at the lattice sites \cite{moreno:prl:2007,likos:cpc:2008,noterods,
matena:pre:2010}. It was found that the separating energy barrier between
neighboring sites scales as $\Delta U \approx 2.3\rho_B$, leading to Arrhenius
behavior $D_B \sim \exp[-\Delta U/T]$ 
\cite{moreno:prl:2007,likos:cpc:2008}. Data of \Fref{Fig:diff} show that this
observation is not altered in the mixture over a broad range of values of $\rho
x/T$. The introduction of the non-clustering $A$-particles does not even change
significantly the former activation energy $\Delta U$.

\subsection{Scattering functions}\label{subsec:scatt}

Relaxation of density fluctuations of wave vector $q$ are evaluated by means of
the intermediate coherent and incoherent scattering functions. The coherent
function is defined as 

\begin{equation}
F_{\alpha\beta}(q,t)=\frac{\langle\rho_\alpha({\bf q},t)\rho_\beta(-{\bf
q},0)\rangle}
{\langle\rho_\alpha({\bf q},0)\rho_\beta(-{\bf q},0)\rangle}
\end{equation}
and characterizes collective $\alpha$-$\beta$ correlations, whereas the
incoherent function accounts for self-correlations and is given by
\begin{equation}
F_{\alpha}^{(\rm s)}(q,t)= \frac{1}{N_\alpha} \left\langle
\sum_{j=1}^{N_\alpha}\exp\{i\mathbf{q}\cdot[\mathbf{r}_{\alpha,j}(t)-\mathbf{r}_
{\alpha,j}(0)]\}\right\rangle.
\end{equation}

\begin{figure}[!t]
\begin{center}
\includegraphics*[width=0.93\textwidth]{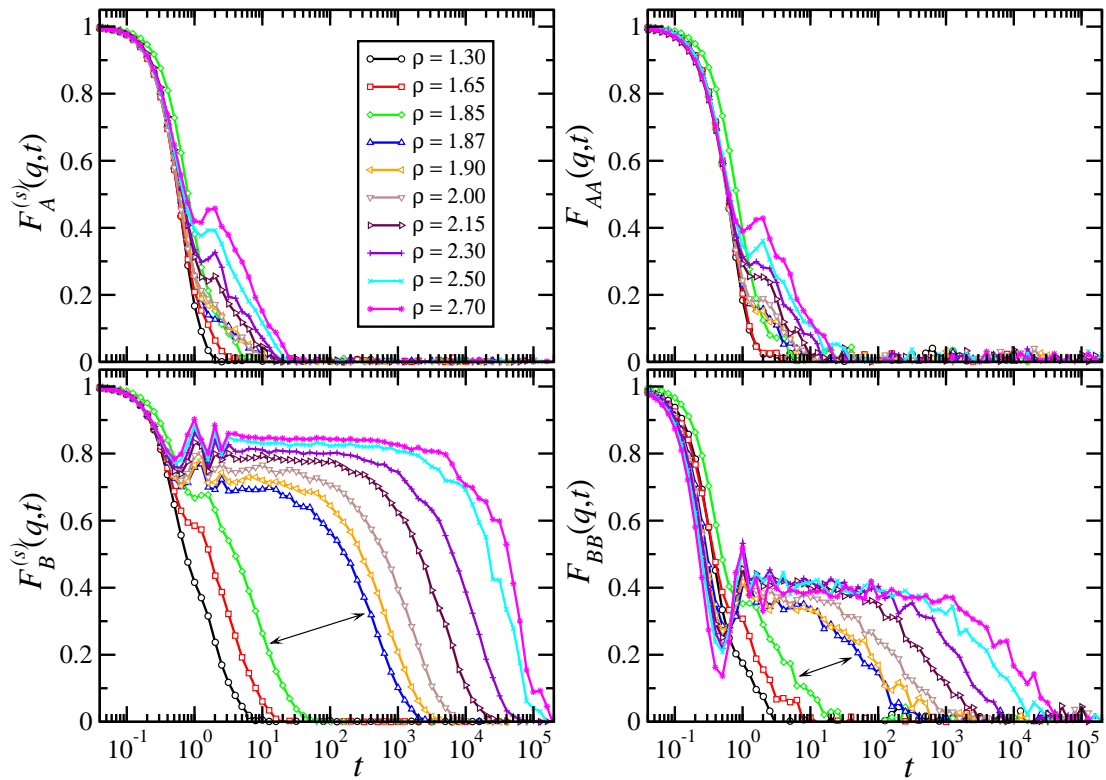}
\end{center}
\caption{Intermediate incoherent and coherent scattering functions for fixed
$x=0.95$ and $q=4.0$, at different densities (see legend for $F_{A}^{\rm
(s)}(q,t)$). The arrows in bottom panels separate fluid states for the
$B$-particles from crystal clusters (see also \Fref{Fig:phase}).} 
\label{Fig:fsqt}
\end{figure}

\begin{figure}[!t]
\begin{center}
\includegraphics*[width=0.93\textwidth]{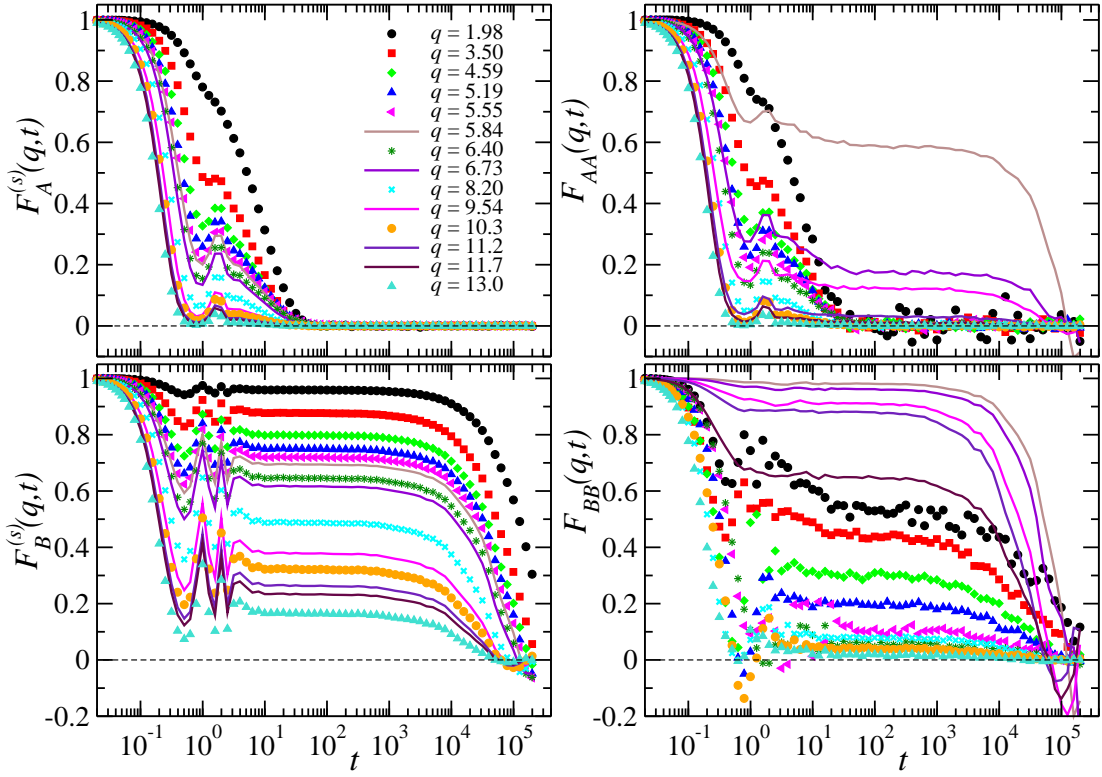}
\end{center}
\caption{Intermediate incoherent and coherent scattering functions for fixed
$x=0.95$ and $\rho=2.70$, at different wave vectors (see legend for $F_{A}^{\rm
(s)}(q,t)$). Solid lines and symbols correspond respectively to RLVs and to
non-RLVs. The dashed lines indicate the zero value of the correlators.}
\label{Fig:fqt-q}
\end{figure}

With the used normalizations 
$F_{\alpha\beta}(q,0)=F_{\alpha}^{(\rm s)}(q,0)= 1$.
\Fref{Fig:fsqt} shows results of
the scattering functions for composition $x = 0.95$ at densities above and below
the crystallization line. The selected value of $q = 4.0$ is not a reciprocal
lattice vector (RLV) and therefore data in this figure reflect relaxation of
out-of-lattice correlations. Consistently with the observations in the real
space (see above), both incoherent and coherent functions for the species $A$
exhibit a fast decay at all the investigated densities, also for those where the
$B$-particles form the cluster crystal. This is reminiscent of the scenario
presented for the small particles in Yukawa mixtures immediately below the
crystallization temperature \cite{kikuchi:epl:2007}. In close similarity to
the aforementioned work,
our data reveal, 
for the crystalline phase, a weak oscillation at $t \sim
2$, immediately after the microscopic decay. It reflects short-time vibrational
dynamics of the $A$-particles within the interstitials. This effect was not
detected in the respective MSD (see above), possibly because contrary to the
scattering functions, the former is generally dominated by fast contributions,
which may mask the oscillations. As discussed above, these short-time
oscillations and the similar ones observed for the scattering functions of the
$B$-particles (Figures \ref{Fig:fsqt} and \ref{Fig:fqt-q}) are artifacts of the
Newtonian dynamics and will be strongly damped or even vanish for BD and MC
simulations.

Concerning the dynamics of the clustering $B$-particles, the present system
exhibits more complex features than usual crystals. As in real-space quantities
(see above), coherent and incoherent functions reflect an abrupt slowing down of
the dynamics by crossing the crystallization point. The intermediate plateau
regime extends over longer time intervals as density increases, before the
ultimate relaxation at long times. This observation is reminiscent of the usual
dynamic scenario associated to the glass transition. Having noted this, distinct
features are revealed by inspection of incoherent and coherent functions. Thus,
the relaxation of the coherent function $F_{BB}(q,t)$ in the cluster crystal
phase is consistent within statistics with a same plateau height (Debye-Waller
factor, $f_q$) for all the investigated densities. This is not the case for the
incoherent function $F_{B}^{\rm (s)}(q,t)$, where the plateau height
(Lamb-M\"{o}ssbauer factor, $f_q^{\rm (s)}$) clearly grows up as density
increases. 

The usual observation in glass-forming systems is that both Debye-Waller and
Lamb-M\"{o}ssbauer factors remain constant on approaching the glass transition
from the ergodic phase. On crossing the transition and moving deep into the
glassy phase, the former factors $f_q$ and $f_q^{\rm (s)}$ progressively
increase. This feature reflects the progressive decrease of the localization
length respectively associated to collective and self-motions. 
With these ideas in mind, the increase of the plateau height in the incoherent
functions for the $B$-particles (\Fref{Fig:fsqt}) suggests that a localization
transition has occurred for the self-motions, tentatively at the
crystallization density. On the contrary, such a transition is not detected for
out-of-lattice collective correlations at any of the investigated crystalline
states. This dynamic decoupling, in the meaning of a different locus for the
localization transitions of self- and out-of-lattice collective motions, is also
present in the pure cluster crystal \cite{moreno:prl:2007,likos:cpc:2008}.
Having noted this, the ergodicity of self-motions is, as in the pure GEM-8
system, restored at long times. This is reflected in the full decay of $F_{\rm
B}^{\rm (s)}(q,t)$. In other words, the localization transition for self-motions
of the $B$-particles is actually avoided. The mechanism which restores
ergodicity of self-motions is naturally the incessant hopping between
neighboring clusters discussed above. Qualitatively similar findings are
observed for the other studied compositions $x =0.8$ and $x = 0.65$ (not shown).
Thus, the addition of non-clustering $A$-particles, even up to a 35\% of the
total,
does not alter the former scenario
of dynamic decoupling in the $B$-particles, and the (avoided by hopping)
localization transition for self-motions still occurs at much lower densities
that for out-of-lattice collective correlations.\footnote{It should be
mentioned,
however, that although the $A$-particles constitute 35\% of the {\it number
density} of the system, they only account for less that 1\% of its 
{\it volume fraction} due to their much smaller size.}

As discussed in Refs. \cite{moreno:prl:2007,likos:cpc:2008}, the former scenario
has similarities and differences with the dynamics in plastic crystals, where
molecules are constrained to vibrate around lattice sites, but perform full
rotations which relax out-of-lattice collective correlations. In the case of
cluster crystals, cluster deformation constitutes an additional mechanism of
relaxation for such correlations. However, contrary to the case of plastic
crystals, self-motions can explore arbitrarily long distances in cluster
crystals
by means of activated hopping. This leads to non-zero diffusivities
(\Fref{Fig:diff}) and full relaxation of incoherent scattering functions
(\Fref{Fig:fsqt}).

As discussed above, the data presented in \Fref{Fig:fsqt} for a selected non-RLV
$q = 4.0$ display similar features, in the case of the species $B$, to those of
the pure GEM-8 system. Novel features are revealed for both species when the
scattering functions are represented as a function of $q$. This is demonstrated
in \Fref{Fig:fqt-q} for a fixed crystalline state point, $x = 0.95$, $\rho =
2.70$, and for different values of $q$, including RLVs and non-RLVs. In the case
of the $A$-particles, the incoherent function displays no signatures of glassy
dynamics, but a fast decay for all the wave vectors. This is also the case for
the coherent function at non-RLVs. Indeed the respective time scales are very
similar for common $q$. However, a strong decoupling between self- and
collective dynamics is observed for RLVs. Contrary to the incoherent case, the
coherent function exhibits a plateau over several decades before relaxation at
long times.

This dynamic decoupling is rather different from that above discussed for the
species $B$ at non-RLVs. On the contrary, it is somewhat reminiscent of the
scenario displayed by the small particles in \textit{disordered} binary mixtures
for certain ranges of composition and size ratio
\cite{moreno:jcp:2006,voigtmann:prl:2009}. In such systems the small particles
move along a channel-like structure, formed by the interstitials of a matrix of
large particles which relaxes in a much longer time scale, or which is even in
the glassy state. As a consequence of the quasistatic arrangement of the
interstitials, collective correlations of the small particles decay very slowly,
despite fast self-motions are performed, allowing for the exploration of large
distances. Thus, decoupling of incoherent and coherent scattering functions of
the small particles is observed in all the range of low and moderate wave
vectors which probe the matrix structure
\cite{moreno:jcp:2006,voigtmann:prl:2009}. As discussed above, for the species
$A$ of the mixtures here investigated, the former decoupling is only observed
for RLVs. This reflects the fact that the non-clustering particles are
preferentially located at the interstitials between the $B$-clusters, as was
seen in the static correlation functions (\Fref{Fig:grsq}), and therefore also
move preferentially between them. Since collective correlations between
interstitial positions are probed by the RLVs, a decoupling between incoherent
and coherent dynamics is observed for such wave vectors, in analogy with the
scenario discussed above for disordered mixtures.

Concerning the scattering functions of the clustering species $B$, slow
relaxation is observed for the incoherent and coherent cases, and both for RLVs
and non-RLVs. As usual, Lamb-M\"{o}ssbauer factors show a monotonic
$q$-dependence, while Debye-Waller factors follow the modulations of the
structure factor. Having noted this, coherent functions reveal non-trivial
differences between RLVs and non-RLVs at the longest times of the simulation
window. Namely, $F_{BB}(q,t)$ exhibits, for RLVs, an oscillation around $t \sim
10^5$. This oscillation is apparently absent for non-RLVs.

\begin{figure}[!t]
\begin{center}
\includegraphics*[width=0.93\textwidth]{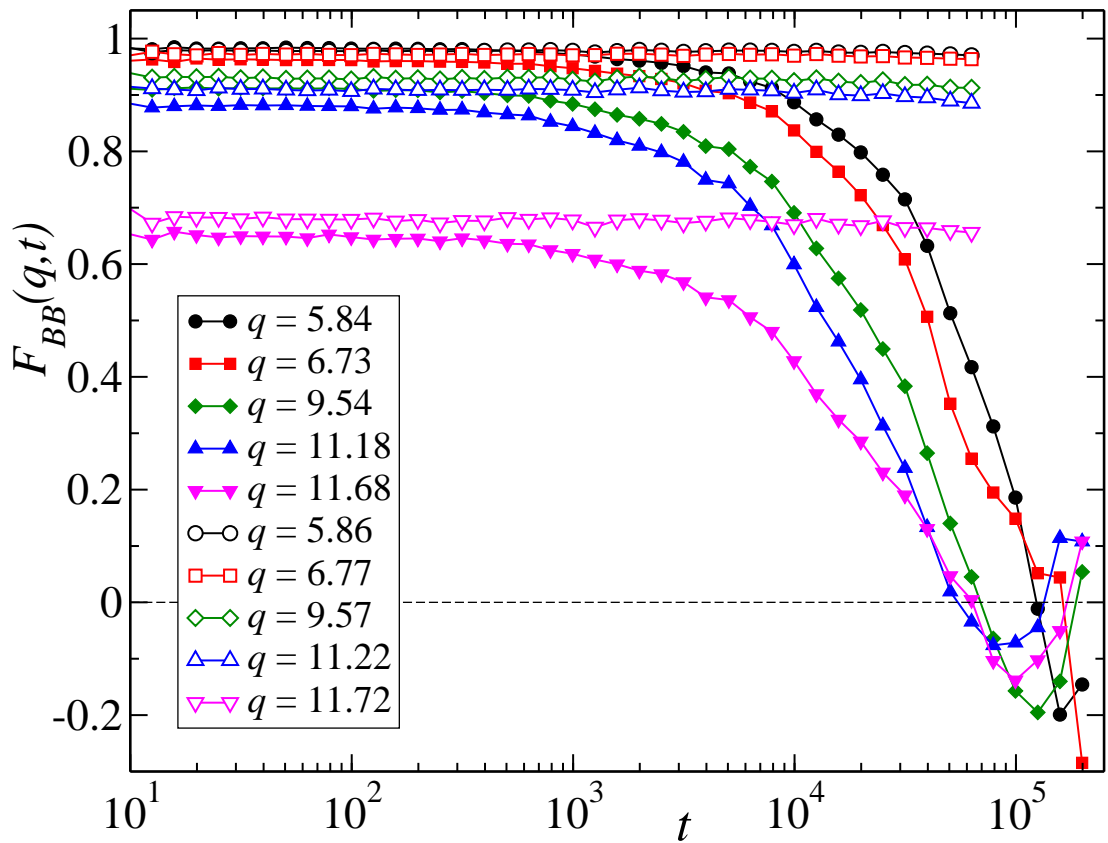}
\end{center}
\caption{Intermediate coherent scattering functions of $B$-particles for several
RLVs (see legend). Filled symbols correspond to the mixture with $x=0.95$,
$\rho=2.70$, and $T = 0.30$. Empty symbols correspond to the pure GEM-8 system
with $\rho = 3.0$ and $T= 0.40$. Solid lines are guides for the eyes. The dashed
line indicates the level $F_{BB}(q,t)=0$.}
\label{Fig:fqt-qlatt}
\end{figure}

For the sake of clarity, the former data for the RLVs are shown separately in
\Fref{Fig:fqt-qlatt} over a time window $t > 10$. Given the large amplitude of
the long-time anticorrelations, reaching values of $F_{BB}(q,t) \cong -0.2$ at
the minimum, it is improbable that the oscillation around $t \sim 10^5$ is an
statistical artifact. The figure also includes coherent data, at its respective
RLVs, of the pure GEM-8 system at $\rho = 3.0$ and $T = 0.40$. The rather
different behavior in the pure system and in the mixture, even with only a 5\%
of $A$-particles in the latter, becomes evident. While full relaxation followed
by oscillation is observed in the mixture, no decay is present in the pure
cluster crystal. It must be stressed that this difference is not related to a
much slower intrinsic dynamics of the pure system for the selected state point
$\rho = 3.0$ and $T = 0.40$. Indeed, by inspection of all the other dynamic
observables introduced above, including coherent functions \textit{for
non-RLVs}, it is found that the pure system shows a faster dynamics than the
mixture at the former state points. For example $D_B \approx 2 \times 10^{-6}$
and $10^{-5}$ respectively for the mixture and the pure system. Therefore, the
oscillatory feature observed at long times in the mixture is induced by the
addition of the gaussian $A$-particles. 

The presence of negative correlations in the collective dynamics of
$B$-particles at near RLV-wavevectors, reached at typical times $\tau_a \sim
10^5$,
is a particular feature of the mixture, which is
is absent for the pure $B$ system within the simulated time window. 
Physically, it describes a process by which a positive density correlation
at $t=0$
at a distance of the order of the lattice constant turns negative at time
scales $\tau_a$. As such, it points to the existence of some characteristic
oscillation frequency $\omega_a \sim \tau_a^{-1}$, which corresponds to
wavevectors lying at the edge of the Brillouin zone, i.e., a short-wavelength
acoustical mode. Although such a process has not been seen for the
pure $B$-system, phonon excitations exist for the pure system as well;
it is thus reasonable to assume that the corresponding frequency
for the pure system is much smaller that $\omega_a$, so that the time
required for this relaxation process to be seen in the coherent scattering
intensities is much longer. A detailed calculation of the phonon spectra
of the mixture at hand, along parallel lines to the one recently carried
out for pure systems \cite{tim:arxiv:2010} would shed light into this
question, it is however cumbersome and it will be left as the subject
of a future publication. Nevertheless, in view of the fact that the
$A$-particles provide additional repulsions to the $B$-species and thus
enhance the restoring forces that act on the latter, it appears plausible
that in the mixture the phonon frequencies are larger than those in the
pure system and thus the characteristic time $\tau_a$ becomes visible
within the simulation window. Alternatively, the issue could also be 
resolved by performing very long simulations for the pure $B$-system.

\section{Conclusions}

The dynamic aspects of binary mixtures of ultrasoft (GEM-8 and gaussian)
particles have been studied by means of MD. The present work extends previous
results for the cluster crystal phase of the pure GEM-8 system, and investigates
the effects of the addition of the non-clustering gaussian particles on the
corresponding dynamic scenario. 
The obtained results show that as the total density increases at fixed
composition, the GEM-8 species builds cluster crystals with a relatively high
localization. As in the one-component system, the incoherent scattering
functions indicate a localization transition for self-motions of the GEM-8
particles, which is avoided by incessant hopping motion between clusters. This
transition seems to occur at lower densities than for out-of-lattice collective
correlations, confirming in the mixture the dynamic decoupling observed in the
one-component cluster crystal.

On the other hand, the gaussian particles remain delocalized in the periodic
potential induced by the GEM-8 clusters and display fast self-motions. However,
slow collective dynamics is observed for specific wave vectors, namely those
belonging to the reciprocal lattice. This feature reflects the preferential
motion
of the gaussian particles over the interstitials, confirming the expectations
from static correlations.
A striking feature is revealed by the analysis of collective
correlations of the GEM-8 particles, for wave vectors at the reciprocal lattice,
a feature which we attribute to the stiffening of an acoustical mode
at the edge of the Brillouin zone, caused by the presence of the $A$-particles.
An additional open question that should be the subject of future investigations
is the influence of the {\it type} of elementary processes employed to
model the dynamics in the mixture, along the lines of the work by
Coslovich {\it et al.}\ carried out for pure 
systems \cite{coslovich:arXiv:2010}. Here,
it should be examined which of the features discovered by means of
our MD approach survive if one employs Monte Carlo moves instead,
which are expected to mimic better the overdamped, Brownian Dynamics
of the system.  

Finally, while there is evidence that both classes of ultrasoft potentials
accurately
model the interactions between real macromolecules in dilute systems, the
validity of these potentials in dense systems has yet to be verified. The
results presented in this paper add to a growing body of work which suggests
that, the search for dense systems where ultrasoft potentials remain valid is a
worthwhile pursuit.   

\ack{This work has been supported by the projects NMP3-CT-2004-502235 (Network
of Excellence SoftComp, EU) and ITN-234810-COMPLOIDS (Marie Curie Training
Network, EU). M.C. acknowledges a PhD fellowship from the German Academic
Exchange Service (DAAD, Germany)
and the ALECOL
Program (Colombia), as well as support from 
Donostia International Physics Center
(Spain).}


\section*{References}
\begin{thebibliography}{10}

\bibitem{louis:prl:2000}  Louis A~A, Bolhuis P~G, Hansen J~P, and Meijer E~J,
2000 \textit{Phys. Rev. Lett.} \textbf{85} 2522
\bibitem{bolhuis:jcp:2001} Bolhuis P~G, Louis A~A, Hansen J~P, and Meijer E~J,
2001 \textit{J. Chem. Phys.} \textbf{114} 4296
\bibitem{likos:jcp:2002} Likos C~N, Rosenfeldt S, Dingenouts N, Ballauff M,
Lindner P, Werner N and  V\"ogtle F, 2002 \textit{J. Chem. Phys.}  \textbf{117}
1869


\bibitem{likos:pre:2001}  Likos C~N, Lang A, Watzlawek M and L\"owen H, 2001
\textit{Phys. Rev. E} \textbf{63} 031206  
\bibitem{glaser:epl:2007} Glaser M~A, Grason G~M, Kamien R~D, Ko{\v s}mrlj A,
Santangelo C~D and Zieherl P, 2007 \textit{Eur. Phys. Lett.} \textbf{78} 46004

\bibitem{mladek:prl:2008}  Mladek B~M, Kahl G and  Likos C~N, 2008 \textit{Phys.
Rev. Lett.} \textbf{100} 028301 

\bibitem{lenz:sm:2009}  Lenz D~A, Blaak R, and  Likos C~N, 2009 \textit{Soft
Matter} \textbf{5} 2905 
\bibitem{narros:sm:2010} Narros A, Moreno A and Likos C~N, 2010 \textit{Soft
Matter} \textbf{6} 2435 

\bibitem{likos:jcp:2007}  Likos C~N, Mladek B~M,  Gottwald D and Kahl G, 2007
\textit{J. Chem. Phys.} \textbf{126} 224502 
\bibitem{mladek.prl.2008} Mladek B~M,  Kahl G and Likos C~N, 2007 \textit{J.
Chem. Phys.} \textbf{126} 224502  


\bibitem{lang:jphys:2000} Lang A, Likos C~N, Watzlawek M, and L\"{o}wen H. 2000
\textit{J. Phys.: Condens. Matter} \textbf{12} 5087

\bibitem{moreno:prl:2007} Moreno A~J and Likos C~N, 2007 \textit{Phys. Rev.
Lett.} \textbf{99} 107801
\bibitem{likos:cpc:2008} Likos C~N, Mladek B~M, Moreno A~J, Gottwald D, and Kahl
G, 2008 \textit{Comput. Phys. Commun.} \textbf{179} 71

\bibitem{sarah:epl:2009}  Overduin S and Likos C~N, 2009 \textit{Eur. Phys.
Lett.} \textbf{85} 26003
\bibitem{sarah:jcp:2009}  Overduin S and Likos C~N, 2009 \textit{J. Chem. Phys.}
\textbf{131} 034902

\bibitem{hansen:tsl:2006} Hansen J~P and McDonald I~R, 2006 \emph{Theory of
simple liquids} (London, Academic Press)

\bibitem{frenkelbook} Frenkel D and Smit B, 1996 \textit{Understanding Molecular
Simulation} (San Diego, Academic Press)

\bibitem{coslovich:arXiv:2010} Coslovich D, Strauss L and Kahl G,
arXiv:1006.4982

\bibitem{noterods} This hopping scenario shows striking analogies, both for van
Hove functions and potential energy profiles, with the layer-to-layer diffusion
of rods in smectic phases (see Ref. \cite{matena:pre:2010}). In such systems the
corresponding energy barriers separate minima located \textit{in the centers} of
the layers. The incoherent scattering functions of the rods and the
$B$-particles (see section \ref{subsec:scatt}) also exhibit similar trends
\cite{matena:pre:2010}.

\bibitem{matena:pre:2010} Matena R, Dijkstra M and Patti A, 2010 \textit{Phys.
Rev. E} \textbf{81} 021704

\bibitem{kikuchi:epl:2007} Kikuchi N and Horbach J, 2007  \textit{Eur. Phys.
Lett.} \textbf{77} 26001

\bibitem{moreno:jcp:2006} Moreno A~J and Colmenero J, 2006 \textit{J. Chem.
Phys.} \textbf{125} 164507

\bibitem{voigtmann:prl:2009} Voigtmann Th and Horbach J, 2009 \textit{Phys. Rev.
Lett.} \textbf{103} 205901

\bibitem{tim:arxiv:2010} Neuhaus T and Likos C N, arXiv:1008.1881


\end{thebibliography}
\end{document}